# Humean Supervenience in the Light of Contemporary Science

VASSILIOS KARAKOSTAS[*]

**Abstract**   It is shown that Lewis' ontological doctrine of Humean supervenience incorporates at its foundation the so-called separability principle of classical physics. In view of the systematic violation of the latter within quantum mechanics, the claim that contemporary physical science may posit non-supervenient relations beyond the spatiotemporal ones is reinforced on a foundational basis concerning constraints on the state-representation of physical systems. Depending on the mode of assignment of states to physical systems — unit state vectors versus statistical density operators — we distinguish between strongly and weakly non-Humean, non-supervenient relations. It is demonstrated that in either case the relations of quantum entanglement constitute prototypical examples of irreducible physical relations that do not supervene upon a spatiotemporal arrangement of Humean qualities, weakening, thereby, the thesis of Humean supervenience. It is examined, in this respect, the status of Lewis' recombination principle, whereas his conception of lawhood is critically investigated. It is concluded that the assumption of ontological reductionism, as expressed in Lewis' Humean doctrine, cannot be regarded as a reliable code of the nature of the physical world and its contents. It is proposed instead that — due to the undeniable existence of non-supervenient relations — a metaphysic of relations of a moderate kind ought to be acknowledged as an indispensable part of our understanding of the natural world at a fundamental level.

## 1  Humean Supervenience

Over the last couple of decades David Lewis, the systematic late philosopher of the 'plurality of worlds', has been defending a metaphysical doctrine he calls 'Humean supervenience'. According to Lewis (1986a, p. ix), Humean supervenience is the doctrine — inspired by Hume, the great denier of necessary connections — that

> ... all there is to the world is a vast mosaic of local matters of particular fact, just one little thing and then another. ... We have geometry: a system of external relations of spatiotemporal distances between points. Maybe points of space-time itself, maybe point-sized bits of matter or aether or fields, maybe both. And at those points we have local qualities: perfectly natural intrinsic properties which need nothing bigger than a point at which to be instantiated. For short: we have an arrangement of qualities. And that is all. There is no difference without difference in the arrangement of qualities. All else supervenes on that.

---

[*] Department of Philosophy and History of Science, University of Athens, Athens 157 71, Greece (E-mail: karakost@phs.uoa.gr)

There is thus a distribution of local, qualitative, intrinsic properties whose instantiation requires no more than a spatiotemporal point.[1] Between these points there exist spatio-temporal relations or occupation relations holding between point-objects and space-time points, or both. These relations do not supervene on the local, intrinsic properties of the related objects. The properties of everything else supervene on the distribution of intrinsic properties instantiated at space-time points or arbitrarily small regions of space-time. The world, therefore, is fragmented into local matters of particular fact and everything else supervenes upon them in conjunction with the spatiotemporal relations among them.

Since the notion of supervenience has been given a multiplicity of various technical definitions in order to help characterize, both in metaphysics and philosophy of mind, a wide variety of philosophical purposes (see, for instance, Savellos and Yalsin 1995; Kim 1993), a clarification of the leading notion of supervenience within Lewis' philosophical framework is required. Lewis defends the usefulness of the notion of supervenience, characterizing supervenience itself as a denial of independent variation. He notes:

> To say that so-and-so supervenes on such-and-such is to say that there can be no difference in respect of so-and-so without difference in respect of such-and-such. Beauty of statues supervenes on their shape, size and colour, for instance, if no two statues, in the same or different worlds, ever differ in beauty without also differing in shape or size or colour (1983, p. 358).

Lewis explicitly accommodates a dependence thesis with his view that "supervenience means that there *could* be no difference of the one sort without difference of the other sort", adding that "without the modality [indicated by 'could'] we have nothing of interest" (1986b, p. 15).

Thus, Lewis' conception of supervenience acquires in effect the form of a 'dependence – determination' relationship. The dependence aspect is that possessing a supervenient (or upper-level) property requires possessing some subvenient (or lower-level) property, whereas, the determination aspect is that possession of that subvenient property will suffice for possession of the supervenient property. Thus, on this conception of supervenience, possessing a supervenient property requires possessing some subvenient property whose possession suffices for the instantiation of the supervenient property in question. In other words, supervenient properties occur only because of the underlying or subvenient properties, and these are sufficient to determine (not necessarily explain) how the supervenient properties arise. It is in this respect that Lewis (1999, p. 29) claims that "a supervenience thesis is, in a broad sense, reductionist".

If, therefore, Lewis' core idea of supervenience is that once the underlying subvenient level is fixed, the upper or supervenient level is fixed as well, then, at least, it should be true that if two objects (e.g., the two statues in Lewis' quotation above) do not differ with respect to their subvenient properties, then they should also not differ with respect to the supervenient properties. Consequently, properties of, say, type-*A* are supervenient on properties of type-*B* if and only if two objects cannot differ with respect to their *A*-properties without also differing with respect to their *B*-properties. Thus, for example, global properties of a compound physical system, considered as a whole, supervene on local properties of its component parts if and only if there can be no relevant difference in the whole without a difference in the parts.

As a 'fairly uncontroversial' example of what supervenience can be like, Lewis (1986b, p. 14) offers the following:

> A dot-matrix picture has global properties — it is symmetrical, it is cluttered, and whatnot — and yet all there is to the picture is dots and non-dots at each point of the matrix. The global properties are nothing but patterns in the dots. They supervene: no two pictures could differ in their global properties without differing, somewhere, in whether there is or isn't a dot.

A dot-matrix pattern is of course supervenient upon the contingent arrangement of dots. The pattern is entailed by the intrinsic properties and distance relations among the dots. If the dots are there, the pattern is also there, in a manner that is entailed by the subvenient base of the dots. No two patterns could differ in their global properties (e.g., symmetry) without differing in their point-by-point arrangement of dots. Lewis, in formulating his thesis of Humean supervenience, takes the view that an analogous kind of entailment relation holds for the totality of all facts about the world, in the sense that all global matters of fact supervene upon a spatiotemporal arrangement of local base facts. In his words:

> Could two worlds differ … without differing, somehow, somewhere, in local qualitative character ? (1986b, p. 14).
> The question turns on an underlying metaphysical issue. A broadly Humean doctrine (something I would very much like to believe if at all possible) holds that all the facts there are about the world are particular facts, or combinations thereof. This need not be taken as a doctrine of analyzability, since some combinations of particular facts cannot be captured in any finite way. It might better be

> taken as a doctrine of supervenience: if two worlds match perfectly in all matters of particular fact, then match perfectly in all other ways too (1986a, p. 111), [specifying that],
>
> the world has its laws of nature, its chances and causal relationships; and yet – perhaps! – all there is to the world is its point-by-point distribution of local qualitative character (1986b, p. 14).

In a world or part of a world where Lewis' doctrine of Humean supervenience holds, there are no necessities in and between the particular facts as such. All the particulars of such an aggregate whole are in their spatiotemporal existence both logically and nomologically independent of each other. Assuming the validity of Lewis' thesis of Humean supervenience, such a contingent spatiotemporal arrangement of local qualities provides an irreducible subvenient basis — what one may call a Humean basis — upon which all else supervenes. The Humean basis consists of intrinsic qualities instantiated locally at space-time points and external distance relations between them. There is no difference anywhere without a difference in the spatiotemporal distribution of intrinsic local qualities. Everything not itself in the Humean basis supervenes upon it. Hence, in Lewis' conception of Humean supervenience, spatiotemporal relations enjoy a privileged status; they are acknowledged as the *only fundamental external physical relations* that are required in the subvenient base. "In a world like ours", Lewis (1994, p. 474) remarks, "the fundamental relations are exactly the spatiotemporal relations: distance relations, both spacelike and timelike … In a world like ours, the fundamental properties are local properties: perfectly natural intrinsic properties of points, or of point-sized occupants of points. … all else supervenes on the spatiotemporal arrangement of local qualities".

In strengthening his conclusion, David Lewis in "On the Plurality of Worlds", while discussing the issue of trans-world comparisons, argues extensively that the *only* external physical relations that may unify a world "have to be, if not strictly spatiotemporal, at least analogically spatiotemporal" (1986b, p. 78). To this purpose, he considers the possibility of whether there could be unifying external relations besides the (strictly or analogically) spatiotemporal interrelations that hold among the parts of a world. He provides, as a test case, the following imaginary scenario:

> We tend to think that positive and negative charge are natural intrinsic properties of particles; but suppose not. Suppose instead there are natural external relations of like-chargedness and opposite-chargedness.... On this view, as opposed to the standard view, the relations of like- and opposite-chargedness do *not* supervene on the intrinsic natures of two particles taken separately; an electron and a positron may be perfect intrinsic duplicates. That is the point of calling the relations external. They are natural *ex hypothesi*. They are pervasive (at least, given the appropriate laws) in that whenever two particles are connected by a chain of such relations, they are connected directly. But they are very far from discriminating (again, given the appropriate laws): if there are as few as three particles, there must be two of them that are alike so far as these relations are concerned. If this story, or something like it, could be true, then here we have external relations that are not strictly or analogically spatiotemporal (1986b, p. 77).

Obviously, Lewis' imaginary relations of like- and opposite-chargedness are not Humean supervenient. For, in such a virtual situation, there simply exist no appropriate intrinsic properties of the related particles instancing the relations in question. Any possible world in which such external relations would hold, it would be a world where Humean supervenience fails. Lewis argues, to this effect, that these imaginary charge relations are unable to unify a world,[2] because there are no plausible reasons against considering particles at *different* worlds also standing in these external relations of like- or opposite-chargedness to each other. However this may be, Lewis has repeatedly conceded that Humean supervenience holds at our actual world. Hence, the main question remains: Are there in our world (or in a possible world like ours) any external non-spatiotemporal, non-Humean-supervenient relations? Lewis offers to his readers the following reply:

> Perhaps there might be extra, irreducible external relations, besides the spatiotemporal ones; there might be emergent natural properties of more-than-point-sized things; there might be things that endure identically through time or space, and trace out loci that cut across all lines of qualitative continuity. It is not, alas, unintelligible that there might be suchlike rubbish. … But if there is suchlike rubbish, say I, then there would have to be extra natural properties or relations that are altogether alien to this world. Within the inner sphere of possibility, from which these alien intrusions are absent, there is indeed no difference of worlds without a difference in their arrangements of qualities (1986a, p. x).

However, as will be shown in the sequel,[3] the conceptualization of the natural world through contemporary physical science — far from abolishing from the inner sphere of possibility as alien any non-spatiotemporal, non-Humean-supervenient relations — strongly suggests the existence of precisely such external irreducible relations.

## 2  Humean Supervenience and its Affinity to Classical Physics

Evidently, Lewis' formulation of Humean supervenience gives expression to some typical Humean claims like the 'looseness' and 'separateness' of things[4] in a manner that is cashed out in terms of intrinsic properties instantiated at distinct space-time points or arbitrarily small regions of space-time. It is important to realise in this respect that Lewis' Humean doctrine incorporates at its foundation the so-called separability principle of classical physics, a principle that essentially matches the notion of atomism that classical physics is assumed to implicitly uphold.[5] Howard (1989, p. 226), drawing inspiration from Einstein's (1948/1971) views on physical reality, offers a formulation of the principle of separability, as a fundamental metaphysical constraint on physical systems and their associated states, along the following lines:

> *Separability Principle.* Each physical system possesses its own distinct, separate state — which determines its qualitative, intrinsic properties — and the whole state of any compound system is completely determined by the separate states of its subsystems and their spatiotemporal relations (cf. Karakostas 2007, p. 280).

Two points of consideration are here in order: Firstly, the aforementioned characterization of the separability principle implies in a natural manner a supervenience claim; for, if the whole state of a compound system is completely determined by the separate states of its subsystems, then the whole state necessarily supervenes on the separate states. Secondly, it is compatible with Lewis' conception of Humean supervenience. As one may recall, Lewis' version of Humean supervenience consists in conceiving the physical properties on which everything else supervenes as properties of points of space-time. In this case, separability about states of physical systems follows straightforwardly. For, if one starts from Humean supervenience, the contents of any two spatiotemporally separated points can be considered to constitute separate physical systems, the joint state of which is fully determined by the local properties that are attributed to each of these points.

Both doctrines of separability and Humean supervenience are inspired by classical physics, be it point-like analytic or field theoretic. With respect to the latter, for instance, the essential characteristic of any classical field theory, regardless of its physical content and assumed mathematical structure, is that the values of fundamental parameters of a field are well defined at every point of the underlying manifold. In the case, for example, of general relativity (*qua* classical field theory), exhaustive specification of the ten independent components of the metric tensor at each point within a given region of the space-time manifold, completely determines the gravitational field in that region. In this sense, the total existence of a field in a given region is contained in its parts, namely, its space-time points. Thus, in attributing physical reality to point-values of basic field parameters, a field theory proceeds by tacitly assuming that a physical state is ascribed to each point of the manifold, and this state determines the local properties of this point-system. Furthermore, the compound state of any set of such point-systems is completely determined by the individual states of its constituents. Hence, classical field theories necessarily satisfy the separability principle and are therefore subjected to the doctrine of Humean supervenience.[6]

Similar considerations arise through the particle-theoretic viewpoint of classical physics. Within the framework of point-like analytic mechanics, the state of a compound system S consisting of N point particles is specified by considering all pairs $\{q_{3N}(t), p_{3N}(t)\}$ of the physical quantities of position and momentum of the individual particles instantiated at distinct space-time points. Hence, at any temporal moment t, the individual pure state of S consists of the N-tuple $\omega = (\omega_1, \omega_2, ... , \omega_N)$, where $\{\omega_i\} = \{q_i, p_i\}$ are the pure states of its constituent subsystems. Consequently, every property the compound system S has at time t, if encoded in $\omega$, is determined by $\{\omega_i\}$. For example, any classical physical quantities (such as mass, momentum, angular momentum, kinetic energy, center of mass motion, gravitational potential energy, etc.) pertaining to the overall system are determined in terms of the corresponding local quantities of its parts. They either constitute direct sums or ordinary functional relations (whose values are well specified at each space-time point) of the relevant quantities of the subsystems. Thus, they are wholly determined by, and hence supervenient on, the subsystem states. In this respect, every classical compound system S is not only admissible to the dictates of the doctrine of Humean supervenience, but, most importantly, it is separable. As such, it is necessarily Humean supervenient upon the joint existence of its relata, namely, its constituent subsystems $S_1, S_2, …, S_N$. Consequently, every qualitative, intrinsic property and relation pertaining to a classical system S, considered as a whole, is supervenient upon the qualitative, intrinsic properties of its basic related parts, in conjunction with their spatio-temporal relations.

What exactly does it mean, however, for a relation to supervene upon the qualitative, intrinsic (and therefore non-relational) properties of the relata to which it refers? Cleland (1984, p. 25) provides a formal characterization of a supervenient relation in modal terms that captures the essential gist of our preceding analysis of supervenience between properties as a 'dependence-determination' relationship:

A dyadic relation *R* is *supervenient* upon a determinable non-relational attribute *P* if and only if:

(1)   $(\forall x, y) \sim \Diamond \{R(x, y)$ and there are no determinate attributes $P_i$ and $P_j$ of determinable kind *P* such that $P_i(x)$ and $P_j(y)\}$;

(2)   $(\forall x, y) \{R(x, y) \supset$ there are determinate attributes $P_i$ and $P_j$ of determinable kind *P* such that $P_i(x)$ and $P_j(y)$ and $(\forall x, y) [(P_i(x)$ and $P_j(y)) \supset R(x, y)]\}$.[7]

The above two conditions can be explicated as follows: If relation *R* is genuinely supervenient upon *P*, then condition (1) implies that *R* cannot possibly appear in the absence of each of its relata instancing the requisite property *P*, whereas condition (2) adds that there must exist one or more pairs of determinate non-relational properties (of kind *P*) whose exemplification alone is sufficient to guarantee the appearance of *R*.

Stating these two necessary and sufficient conditions as a characterization of supervenience, enables Cleland to distinguish between two kinds of 'non-supervenience' in terms of strongly and weakly non-supervenient relations:

*Strong non-supervenience*: a relation is strongly non-supervenient if the appearance of this relation is neither dependent upon nor determined by non-relational properties of its relata. In other words, a relation is strongly non-supervenient if and only if it violates both conditions (1) and (2). That is to say, if two entities bear a strongly non-supervenient relation to each other, then they necessarily should possess no intrinsic properties of a certain kind instancing the relation in question.

*Weak non-supervenience*: a relation is weakly non-supervenient if the appearance of this relation is dependent upon the instantiation of non-relational properties of each of its relata, but the latter are not sufficient to determine the relation in question. Accordingly, a relation is weakly non-supervenient if and only if it satisfies condition (1), but violates condition (2).

It should be observed that Cleland's formulation of a supervenient relation and its negation is independent of any predilections concerning the conceptual foundations of contemporary physics. On the other hand, French (1989), inspired by Teller (1986), introduced Cleland's conditions into the interpretation of quantum theory as a reasonable alternative to understanding the puzzling EPR-entangled correlations in terms of non-supervenient relations. In the next section, Cleland's conditions are projected into the context of quantum nonseparability as a means of rigorously evaluating the compatibility status of Humean supervenience against the background of foundational physics.

**3 Quantum Entangled Relations and Humean Supervenience**

Classical physics, as already demonstrated, provides strong motivation to Lewis' thesis of Humean supervenience, at least, to the extent that the latter incorporates at its foundation the aforementioned separability principle of Section 2. The notion of separability has been regarded within the framework of classical physics as a principal condition of our conception of the world, a condition that characterizes all our thinking in acknowledging the physical identity of distant things, the "mutually independent existence (the 'being thus')" of spatiotemporally separated systems (Einstein 1948/1971, p. 169). The principle of separability delimits in essence the fact — upon which the whole classical physics is implicitly founded — that any compound physical system of a classical universe can be conceived of as consisting of *separable*, *individual* parts interacting by means of forces, which are encoded in the Hamiltonian function of the overall system, and that the state of the latter (in the limiting case, the physical state of the world itself) is completely determined by the intrinsic physical properties pertaining to each one of these parts and their spatiotemporal relations. In contradistinction, standard quantum mechanics systematically violates the conception of separability that classical physics accustomed us to consider as valid.[8] From a formal point of view, the source of its defiance is due to the tensor-product structure of a compound Hilbert space and the quantum mechanical principle of the superposition of states, incorporating a kind of objective indefiniteness for the numerical values of any physical quantity belonging to a superposed state.

*3.1 Strongly Non-Humean Non-Supervenient Relations*

As a means of explicating the preceding factors in relation to Lewis' doctrine of Humean supervenience, let us consider the simplest possible case of a compound system S consisting of a pair of subsystems $S_1$ and $S_2$ with corresponding Hilbert spaces $H_1$ and $H_2$.

Naturally, subsystems $S_1$ and $S_2$, in forming system S, have interacted by means of forces at some time $t_o$ and suppose that at times $t > t_o$ they are spatially separated. Then, any pure state W of the compound system S can be expressed in the tensor-product Hilbert space $H = H_1 \otimes H_2$ in the Schmidt form

$$W = P_{|\Psi>} = |\Psi><\Psi| = \sum_i c_i (|\psi_i> \otimes |\varphi_i>), \qquad \| |\Psi> \|^2 = \sum_i |c_i|^2 = 1 , \qquad (1)$$

where $\{|\psi_i>\}$ and $\{|\varphi_i>\}$ are orthonormal vector bases in $H_1$ (of $S_1$) and $H_2$ (of $S_2$), respectively.

If there is just one term in the W-representation of Eq. (1), i.e., if $|c_i| = 1$, the state $W = |\psi> \otimes |\varphi>$ of the compound system forms a product state: a state that can always be decomposed into a single tensor-product of an $S_1$-state and an $S_2$-state. In this circumstance, each subsystem of the compound system possesses a separable and well-defined state, so that the state of the overall system consists of nothing but the logical sum of the subsystem states in consonance with the separability principle of Section 2. This is the only highly particular as well as idealised case in which a separability principle holds in quantum mechanics. For, even if a compound system at a given temporal instant t is appropriately described by a product state — $W(t) = |\psi_{(t)}> \otimes |\varphi_{(t)}>$ — the preservation of its identity under the system's natural time evolution — $W(t_2) = U(t_2-t_1) W(t_1)$, for all $t \in R$ — implies that the Hamiltonian $H$ (i.e., the energy operator of the system) should be decomposed into the direct sum of the subsystem Hamiltonians — $H = H_1 \otimes I_2 + I_1 \otimes H_2$ — and this is precisely the condition of *no interaction* between $S_1$ and $S_2$ (e.g., Blank et al. 1994, ch. 11). Obviously, in such a case, subsystems $S_1$ and $S_2$ behave in an entirely uncorrelated and independent manner. Correlations, even of a probabilistic nature, among any physical quantities corresponding to the two subsystems are simply non existent, since for any two observables $A_1$ and $A_2$ pertaining to $S_1$ and $S_2$, respectively, the probability distributions of $A_1$ and of $A_2$ are disconnected: Tr $(A_1 \otimes A_2)$ $(|\psi> \otimes |\varphi>) = $ Tr $(A_1 |\psi>) \cdot$ Tr $(A_2 |\varphi>)$.

If, however, there appear more than one term in the W-representation of the compound system, i.e., if $|c_i| < 1$, then there exist entangled correlations (of the well-known EPR-type) between subsystems $S_1$ and $S_2$. It can be shown in this case, as already indicated by Schrödinger (1935/1983, p. 161), that there are no subsystem states $|\xi>$ ($\forall$ $|\xi> \in H_1$) and $|\chi>$ ($\forall$ $|\chi> \in H_2$) such that W is equivalent to the conjoined attribution of $|\xi>$ to subsystem $S_1$ and $|\chi>$ to subsystem $S_2$, i.e., $W \neq |\xi> \otimes |\chi>$. Thus, when a compound system, such as S, is in an entangled state W, namely a superposition of pure states of tensor-product forms, neither subsystem $S_1$ by itself nor subsystem $S_2$ by itself is associated with an individual pure state. The normalised unit vectors $|\psi_i\rangle$, $|\varphi_i\rangle$ belonging to the Hilbert space of either subsystem are not eigenstates of the overall state W. If descriptions of physical systems are restricted to the state vector assignment of states, then, strictly speaking, subsystems $S_1$ and $S_2$ have no states at all, even when $S_1$ and $S_2$ are spatially separated. Only the compound system is assigned a definite (nonseparable) pure state W, represented appropriately by a state vector $|\Psi\rangle$ in the tensor-product Hilbert space of S. Maximal determination of the whole system, therefore, does not allow the possibility of acquiring maximal specification of its component parts, a circumstance with no precedence in classical physics.

Since on the state vector ascription of states, neither subsystem $S_1$ nor subsystem $S_2$ has a state vector in S, it is apparent that the state W of the compound system cannot be reproduced on the basis that neither part has a well-defined individual state, namely, a pure state. Thus, the separability principle of Section 2 is violated and likewise Lewis' version of Humean supervenience does fail. The entangled state W represents global properties for the whole system S that are neither *dependent* upon nor *determined* by any properties of its parts. In fact, in any case of quantum entanglement, conceived as a relation among the constitutive parts of a quantum whole, there exist no qualitative, non-relational properties of the parts whose exemplification is sufficient to guarantee the appearance of entanglement. In such a case both supervenience conditions (1) and (2) are clearly violated; for not only the entangled relation pertaining to compound system S is not guaranteed by the exemplification of qualitative, non-relational properties of subsystems $S_1$ and $S_2$, thus violating supervenience condition (2), but neither of the subsystems $S_1$ and $S_2$, being the related parts of the entangled relation, exemplify any qualitative, non-relational properties at all, thus violating supervenience condition (1). Hence, it may be asserted that the relation of quantum entanglement is *inherent* to the compound system S, as a whole, exhibiting strongly non-supervenient behaviour with respect to the relata $S_1$ and $S_2$. In this respect, Lewis' thesis of Humean supervenience fails on, at least, two counts: firstly, there exist non-supervenient relations beyond the spatiotemporal ones, namely quantum mechanical entangled relations, and, secondly, in considering any such relation, there exist no non-relational states (or properties) of its related parts.

As a means of illustrating the preceding points, let us consider an important class of compound systems that form the prototype of quantum entangled systems, namely, spin-singlet pairs. Let S be a compound system consisting of a pair ($S_1$, $S_2$) of spin-1/2 particles in the following EPR-type superposed state, known as the *singlet* state

$$W_S = 1/\sqrt{2}\ \{|\psi_+\rangle_1 \otimes |\varphi_-\rangle_2\ -\ |\psi_-\rangle_1 \otimes |\varphi_+\rangle_2\}, \qquad (2)$$

where $\{|\psi_\pm\rangle_1\}$ and $\{|\varphi_\pm\rangle_2\}$ are spin-orthonormal bases in the two-dimensional Hilbert spaces $H_1$ and $H_2$ associated with $S_1$ and $S_2$, respectively.[9] As is well known, in such a case, it is quantum mechanically predicted and experimentally confirmed that the spin components of $S_1$ and of $S_2$ have always opposite spin orientations; they are perfectly anti-correlated. Whenever the spin component along a given direction of, say, particle $S_1$ is measured at time $t_o$ and found equal to $+1/2\ \hbar$ (correspondingly $-1/2\ \hbar$), the subsequent destruction of the superposition bonds (between the tensor-product states involved) imparts to particle $S_2$ a tendency: that of inducing — in this special case, with certainty — the opposite value $-1/2\ \hbar$ (correspondingly $+1/2\ \hbar$), if and when, at a time $t>t_o$, particle $S_2$ is submitted to an appropriate measurement of the same component of spin as $S_1$. From a physical point of view, this derives from the interference (the anti-symmetric phase interrelations) with which the subsystem unit vectors $|\psi_\pm\rangle_1$ and $|\varphi_\pm\rangle_2$ — or, more precisely, the two product states $|\psi_+\rangle_1 \otimes |\varphi_-\rangle_2$, $|\psi_-\rangle_1 \otimes |\varphi_+\rangle_2$ — are combined within $W_S$. This, in turn, leads not only to the subsystem interdependence of the type described above, but also to conservation of the total angular momentum for the pair ($S_1$, $S_2$) of spin-1/2 particles, and thus to the property of definite total spin of value zero for the compound system S.

The latter is an *irreducible, holistic* property of S: it is not determined by any physical properties of its subsystems $S_1$, $S_2$ considered individually. Specifically, the property of S 'having total spin zero' is strongly non-supervenient on the spin properties of $S_1$ and of $S_2$, since neither $S_1$ nor $S_2$ has any definite spin in the singlet state $W_S$. Moreover, the probability distributions concerning spin components of $S_1$ and of $S_2$ along some one direction do not ensure, with probability one, S's possession of this property. Neither the latter could be understood or accounted for by the possibility (that an adherent of Humean supervenience may favour) of treating $S_1$ and $S_2$ separately at the expense of postulating a relation between them as to the effect of their spin components 'being perfectly anti-correlated'. For, while 'having total spin zero' is an intrinsic physical property of the compound system S in the nonseparable state $W_S$, the assumed relation is not an intrinsic physical relation that $S_1$ and $S_2$ may have in and of themselves. That is, although the relation of perfect anti-correlation is encoded within state $W_S$, ascribing this relation to individual parts of a system is not tantamount to being in state $W_S$. The relation of perfect anti-correlation is inherent to the entangled state $W_S$ itself which refers directly to the whole system. The entangled correlations between $S_1$ and $S_2$ just do not supervene upon any (intrinsic or extrinsic) properties of the subsystem parts taken separately.

It may seem odd to consider non-supervenient relations holding between non-individuatable relata. However, the important point to be noticed is that within an entangled quantum system there is no individual pure state for a component subsystem alone. Within $W_S$ neither subsystem $S_1$ nor subsystem $S_2$ acquire individual independent existence. In considering any entangled compound system, the nature and properties of component parts may only be determined from their 'role' — the forming pattern of the inseparable web of relations — within the whole. Here, the part-whole relationship appears as complementary: the part is made 'manifest' through the whole, while the whole can only be 'inferred' via the interdependent behaviour of its parts. Thus, in the example under consideration, the property of total spin of the whole in the singlet state $W_S$ does indicate the way in which the parts are related with respect to spin, although neither part possesses a definite numerical value of spin in any direction in distinction from the other one. And it is *only* the property of the total spin of the whole that contains *all* that can be said about the spin properties of the parts, because it is only the entangled state of the whole that contains the correlations among the spin probability distributions pertaining to the parts.[10] Consequently, the part-whole reduction with respect to the property of total spin zero in $W_S$ has failed: the latter property, whereas characterizes the whole system, is not supervenient upon — let alone being reducible to — any properties of its constituent parts. Exactly the same holds for the properties of total momentum and relative distance of the overall system S with respect to corresponding local properties of the parts. Analogous considerations, of course, to the aforementioned paradigmatic case of the spin-singlet pair of particles apply to any case of quantum entanglement. Entanglement need not be of maximal anticorrelation, as in the example of the singlet state.[11] It does neither have to be confined to states of quantum systems of the same kind; entanglement reaches in principle the states of all compound quantum systems.

The generic phenomenon of quantum entanglement and the associated conception of quantum nonseparability cast severe doubts on the applicability of the doctrine of Humean supervenience. In fact, the non-supervenient relations of entanglement among the parts of a quantum whole imply a reversal of Lewis' thesis of Humean supervenience: for, there exist properties pertaining to any entangled quantum system which, in a clearly specifiable sense, characterize the whole system but are neither supervenient upon nor reducible to or derived from any combination of local properties of its parts. On the contrary, it is only the entangled state of the whole system which completely determines the local properties of its subsystem parts and their relations (to the extent that these are determined at all; see further Section 3.2). If Lewis' version of Humean supervenience held within quantum mechanics, one could analyse the entangled state of a compound system into local physical states of the component parts taken separately in conjunction with the spatiotemporal relations among the parts. In such a case, however, the state of the compound system would be a product state, in flagrant contradiction with the initial assumption of entanglement. Evidently, Lewis' thesis of Humean supervenience — proclaiming that the spatiotemporal arrangement of local intrinsic qualities provides a subvenient basis upon which all else supervenes — comes into conflict with contemporary physics.[12] Within the framework of quantum mechanics, even the core presupposition of Lewis' Humean assumption that there exist local entities furnishing the bearers of the fundamental intrinsic properties and relations comes into question. For the nonseparable character of the behaviour of an entangled quantum system precludes in a novel way the possibility of describing its component subsystems as well-defined individuals, each with its own pure state or pre-determined physical properties. Upon any case of quantum entanglement, it is not permissible to consider the parts of a quantum whole as self-autonomous, intrinsically defined individual entities. In fact, as considered immediately below, whenever the pure entangled state of a compound system is decomposed in order to represent well-defined subsystems, the effect can only extent up to a representation in terms of statistical (reduced) states of those subsystems.

*3.2 Weakly Non-Humean Non-Supervenient Relations*

In view, therefore, of the radical violation of Humean supervenience on the state vector ascription of quantum states, it is interesting to inquire whether the assignment of statistical states to component subsystems, represented by non idempotent density operators, restores a notion of Humean supervenience into quantum theory? The question, however, even according to such a circumstance of physical possibility, is answered strictly in the negative.

The clearest way to establish this, for present purposes, is by regarding again as the state W of a compound system the singlet state of a pair of spin-1/2 particles ($S_1$, $S_2$) in the familiar development

$$W_S = 1/\sqrt{2} \ \{|\psi_+\rangle_1 \otimes |\varphi_-\rangle_2 - |\psi_-\rangle_1 \otimes |\varphi_+\rangle_2\}. \tag{3}$$

Observe, in consonance with the considerations of Section 3.1, that neither particle $S_1$ nor particle $S_2$ can be represented in $W_S$ of Eq. (3) by a state vector. However, each particle may be assigned a state, albeit a reduced state, that is given by the partial trace of the density operator $W_S$ of the compound system. Recall that the reduced state of each particle arises by 'tracing over' the degrees of freedom associated with the Hilbert space representation of the partner particle. Hence, the following density operators

$$\hat{W}_1 = 1/2 \ P_{|\psi_+\rangle} + 1/2 \ P_{|\psi_-\rangle} \quad \text{and} \quad \hat{W}_2 = 1/2 \ P_{|\varphi_+\rangle} + 1/2 \ P_{|\varphi_-\rangle} \tag{4}$$

represent the reduced ('unpolarized') states of spin-1/2 particles $S_1$ and $S_2$, respectively, in state $W_S$.[13]

It is not hard to show, however, that the component states $\hat{W}_1$ and $\hat{W}_2$ of (4) could be identical if derived from a compound state $W'$ that would correspond to the triplet state

$$W' = 1/\sqrt{2} \ \{|\psi_+\rangle \otimes |\varphi_-\rangle + |\psi_-\rangle \otimes |\varphi_+\rangle\}, \tag{5}$$

or to the following pure states (proviso the sign)

$$W'' = 1/\sqrt{2} \ \{|\psi_+\rangle \otimes |\varphi_+\rangle \pm |\psi_-\rangle \otimes |\varphi_-\rangle\} \tag{6}$$

that yield, in general, different predictions than W or W' does for certain spin measurements of both $S_1$ and $S_2$ along a given direction; or they still could be identical if derived from the mixed state

$$W''' = 1/2 \ \{|\psi_+\rangle \otimes |\varphi_-\rangle + |\psi_-\rangle \otimes |\varphi_+\rangle\}. \tag{7}$$

Thus, given the states $\hat{W}_1$ and $\hat{W}_2$ of subsystems $S_1$ and $S_2$, respectively, the compound state could equally well be either W or W', W'', W''' or in numerous other variations of them. There exists a many-to-one mapping of the subsystem non-pure states to the state of the whole system.[14] Accordingly, the specification of the compound system remains indefinite. For, the compound state $W_S$ contains correlations between subsystems $S_1$ and $S_2$ that the reduced states $\hat{W}_1$ and $\hat{W}_2$ do not contain. The sort of correlations that is missing corresponds, from a formal point of view, to the tracing out in the specification, for instance, of $\hat{W}_1$ of what might be known about the state of subsystem $S_2$ and about its connection with subsystem $S_1$. It is evident, therefore, that at the level of description in terms of a reduced state, the information obtained by considering, even simultaneously, the two subsystems does not permit the reconstruction of the pure state of the whole system, i.e., W≠$\hat{W}_1 \otimes \hat{W}_2$. In this sense, it may be said that the 'whole' (e.g., system S) is, in a non-trivial way, more than the combination of its 'parts' (e.g., subsystems $S_1$ and $S_2$), this being the case even when these parts occupy remote regions of space however far apart. Hence, on the density operator ascription of states, whereas each subsystem possesses a local state independently of the other, still the state of the whole system fails to be exhausted by a specification of the states of the parts and their spatiotemporal relations. Consequently, the separability principle of Section 2 is violated.

If the preceding state of affairs be translated in terms of state-dependent properties, it immediately follows that the exemplification of non-relational, qualitative properties pertaining to $\hat{W}_1$ and $\hat{W}_2$ is not sufficient to guarantee the exemplification of the relational properties incorporated in $W_S$. In such a circumstance, supervenience condition (1) is satisfied, whereas supervenience condition (2) is clearly violated. Thus, even on the statistical assignment of states to component subsystems of a quantum whole, one is still confronted with weakly non-supervenient relations.

The difficulty facing Lewis' doctrine of Humean supervenience is now unavoidable. Consider two pairs of spin-1/2 particles, one in the singlet and the other in the triplet state, represented by Eqs. (3) and (5), respectively, such that the spatiotemporal relations within each pair are identical. Each particle of either pair in the singlet or triplet state can now be assigned its own local state, a reduced statistical state in accordance with the expressions of Eq. (4). It is important to note that both singlet and triplet states assign the same spin probability distributions for each component particle. No local spin-measurements performed on a component particle of a pair in the singlet state can distinguish it from a component of a pair in the triplet. For, the statistical states $\hat{W}_1$ and $\hat{W}_2$ assigned to component particles in the singlet state are *identical* to the statistical states assigned to component particles in the triplet state; they are equiprobable 'mixtures' of the states $|\psi_\pm\rangle$ and $|\varphi_\pm\rangle$. Consequently, if Lewis' version of Humean supervenience holds, then, since each component particle in the singlet state is in precisely the same spin state as each component in the triplet state, and since the spatio-temporal relations between the component particles of each pair are identical, the singlet state would have to be identical to the triplet state in Lewis' terms. These are, however, significantly different. The singlet state gives rise to a global property for total spin of eigenvalue zero, whereas, the triplet state predicts with certainty the eigenvalue result $2\hbar^2$. Both values may be verified by joint measurements on the two-component system for the total (squared) spin operator in the corresponding states. Thus, there is a difference in global properties to which no difference in the local properties of the component parts corresponds. Hence, Lewis' thesis of Humean supervenience fails.

## 4  The Dubious Status of the Recombination Principle

Likewise fails his recombination principle. The latter constitutes a re-expression of the Humean denial of necessary connections between distinct existences. In short, Lewis' recombination principle requires that ''anything can *coexist* with anything, and which thereby prohibits a necessary connection between the intrinsic character of a thing and the intrinsic character of distinct things with which it coexists'' (Lewis 1986b, p. 181; see also pp. 86-92). To realise the scope of Lewis' recombination principle, it is useful to recall that his metaphysical doctrine of Humean supervenience conceives the world as consisting of an enormous number of local matters of particular fact. The world is composed of *discrete* events, like a vast mosaic, ''just one little thing and then another'' with no tie or connection between one and the other. Each event is self-contained, all its basic properties are intrinsic; its nature is exhausted within the spatiotemporal boundary that occupies. Accordingly, each event may coexist alongside with any other particular or particulars to form any pattern. Yet, in Lewis' vision of the world, even if a sequence or a combination of events presents a recognizable pattern, the individual event has a nature and existence *independent* of the pattern they formed. Each event of this pattern is causally inert,[15] thus incapable of reaching out beyond itself to any other event. It is a particular whose nature is completely independent of anything around it. Hence, there are *no* necessary connections between an individual event and any of its neighbours or other events in the pattern. The latter simply consists of the mereological sum of all particulars that are part of it. The same events — by virtue of their mutual independent existence — could have occurred in a different order and arrangement so as to form a completely different pattern. At any circumstance, the events, and their spatiotemporal relations among each other, constitute the whole of the objectively real world. From the perspective of Lewis' metaphysic, if one is able to determine the intrinsic qualities of particular events or atomic objects in space and time, then one can describe the world completely.

Quantum mechanics, however, is not in conformity with Lewis' atomistic metaphysical picture that depicts a world of self-contained, unconnected particulars that exist independently of each other. As we have extensively argued, within the quantum theoretical context, the consideration of physical reality cannot be comprehended as the sum of its parts in conjunction with the spatiotemporal relations among the parts, since the quantum whole provides the framework for the existence of the parts. In considering any case of quantum entanglement, the interrelation between the parts cannot possibly be disclosed in an analysis of the parts that takes no account of the entangled connection of the whole. As already shown, their entangled relation does not supervene upon any intrinsic or relational properties of the parts taken separately. This is indeed the feature which makes the quantum theory go beyond any mechanistic or atomistic thinking. In consistency with Lewis' view, given any compound physical system, the intrinsic properties of the whole ought to be reducible to or supervenient upon the properties of its parts and their spatiotemporal relations. In quantum mechanics the situation is actually reversed; due to the genuinely nonseparable structure of quantum theory, the state-dependent properties of the parts can ultimately be accounted only in terms of the characteristics of the whole.[16] In a truly nonseparable physical system, as in an entangled quantum system, the part does acquire a different identification within the whole from what it does outside the whole, in its own 'isolated', separate state (see esp. Section 3.1). Thus, for instance, no definite spin property of an isolated spin-1/2 particle, e.g. a 'free' or 'bare' electron, can be identified with the spin property of either member of a pair of electrons in the singlet state, since in this situation any spin state can be specified only at the level of the overall system. When in the singlet state, there is simply no individual spin state for a component particle alone, unless explicit reference is made to the partner particle via the total information contained in the compound state. Consequently, any spin state of either particle is fixed only through the interconnected web of entangled relations among the particles. Hence, the spin property of either particle, when in an entangled state, cannot stand alone, unconnected to the whole pattern and unaffected by anything else, in flagrant contradiction with Lewis' suggestion concerning his recombination principle.

## 5  Adjusting Humean Supervenience?

Lewis (1986a, pp. x-xi) acknowledges that quantum theory may pose a threat to Humean supervenience:

> … it just might be that Humean supervenience is true, but our best physics is dead wrong in its inventory of the qualities. Maybe, but I doubt it. … what I uphold is not so much the truth of Humean supervenience as the *tenability* of it. If physics itself were to teach me that it is false, I wouldn't grieve. That might happen: maybe the lesson of Bell's theorem is exactly that there are *physical* entities which are unlocalized, and which might therefore make a difference between worlds … that match perfectly in their arrangements of local qualities. Maybe so. I'm ready to believe it. But I am not ready to take lessons in ontology from quantum physics as it now is. First I must see how it looks when it is purified of instrumentalist frivolity, and dares to say something not just about pointer readings but about the constitution of the world; and when it is purified of doublethinking deviant logic; and — most all — when it is purified of supernatural tales about the power of the observant mind to make things jump. If, after all that, it still teaches nonlocality, I shall submit willingly to the best of authority.

Lewis (1994, p. 474) in addition offers the following qualification regarding his defence of the thesis of Humean supervenience:

> The point of defending Humean Supervenience is not to support reactionary physics, but rather to resist philosophical arguments that there are more things in heaven and earth than physics has dreamt of. Therefore if I defend the *philosophical* tenability of Humean Supervenience, that defence can doubtless be adapted to whatever better supervenience thesis may emerge from better physics.

It should be observed in relation to Lewis' first quote that quantum theory can no longer be viewed as an awkward and annoying phase through which physical science is progressing, only later surely to settle down as 'philosophically more respectable'. Quantum theory has reached a degree of maturity that simply cannot be ignored on the basis of a priori philosophical expectations. In the course of the multifaceted investigations of the conceptual foundations of quantum mechanics, there has been developed a variety of formulations of the theory in which neither the concept of 'measurement' nor the concept of 'the observant mind' bear any fundamental role. Approaches of this kind involve, for instance, the consistent histories formulation of Griffiths (1984) and Omnès (1992), the decoherence theory of Zurek and Gell-Mann (Joos et al. 2003, for a recent review), the stochastic localization model of Ghirardi, Rimini, Weber (1986) and Pearle (1989), or even Everett's (1957) relative-state formulation. All these theories posit fundamental nonseparable physical states of affairs. Nonseparability constitutes a structural feature of quantum mechanics that distinctly marks its entire departure from classical lines of thought. If, therefore, present day quantum theory, or any extension of it, is part of a true description of the world, then Lewis' doctrine of Humean supervenience cannot be regarded as a reliable code of the nature of the physical world and its contents. Quantum theory implies in fact, as extensively argued, an irrevocable failure of Humean supervenience. For, it shows that in considering any compound quantum system, there simply exist no intrinsic local properties of the entangled systems involved which could form a basis on which the relation of entanglement could supervene. Acknowledging the generic character of the latter at the microphysical level, no proper foundation can be established within contemporary physics for formulating the doctrine of Humean supervenience. Consequently, Lewis' qualification that his defence of Humean supervenience "can doubtless be adapted to whatever better supervenience thesis may emerge from better physics", is clearly not the case.[17] The thesis of Humean supervenience is ill-formed; it relies upon a false presupposition, namely separability.

Neither the violation of the separability principle, as established by quantum theory, revives the existence of "more things in heaven and earth than physics has dreamt of ". On the contrary, quantum nonseparability only implies that there are more fundamental relations dictated by physics — namely, quantum mechanical entangled relations — than Humean supervenience can possibly accommodate. The physical existence of quantum entangled relations, as well as their empirical confirmation,[18] undeniably shows that what there is in the world is more tightly intertwined than just by spatiotemporal relations among separately existing entities that are localized at space-time points. In contrast to the thesis of Humean supervenience, therefore, relations of quantum entanglement are proven to be at least as fundamental as spatiotemporal relations in unifying a world.

## 6  Problems for Lewis' Best-System Analysis of Law

No doubt, Lewis' overall metaphysical system is impressive both in its extent and depth. It incorporates a variety of issues involving, for instance, the analysis of concepts, such as causation, truth, counterfactuals, as well as mental and semantic concepts, such as belief and reference, in terms of a sequence of supervenience claims for these concepts. We shall take no stance on these issues, as they fall outside the scope of this paper. In evaluating Humean supervenience from the viewpoint of contemporary science, we shall only briefly dwell on Lewis' Humean account of laws of nature. Namely, Lewis' conception that laws themselves are supervenient upon local particular Humean facts. One of the leading philosophical outlooks on Humean analysis of lawhood is David Lewis' 'best-system' account. Concisely, in this view, laws of nature are understood as the theorems or axioms of the best possible systematization of the world's history of events or facts, where the criterion employed for being the best such systematization is that which accomplishes the best *balance* between *simplicity* and *strength* (i.e., informativeness). Here is how Lewis (1973, p. 73) expresses the claim:

> a contingent generalization is a *law of nature* if and only if it appears as a theorem (or axiom) in each of the true deductive systems that achieves a best combination of simplicity and strength. A generalization is a law at a world *i*, likewise, if and only if it appears as a theorem in each of the best deductive systems true at *i*.

And again:

> Take all deductive systems whose theorems are true. Some are simpler, better systematized than others. Some are stronger, more informative than others. These virtues compete: an uninformative system can be very simple, an unsystematized compendium of miscellaneous information can be very informative. The best system is the one that strikes as good a balance as truth will allow between simplicity and strength. … A regularity is a law iff it is a theorem of the best system. (Lewis 1994, p. 478).

To account also for the case of probabilistic laws, Lewis introduces into the preceding best-system analysis of laws the notion of chance. Then, in his definition of law, he replaces the set of all true deductive systems by the set of those systems which never had any chance of being false. He writes:

> … we modify the best-system analysis to make it deliver the chances and the laws that govern them in one package deal. Consider deductive systems that pertain not only to what happens in history, but also to what the chances are of various outcomes in various situations … Require these systems to be true in what they say about history. We cannot yet require them to be true in what they say about chance, because we have yet to say what chance means; our systems are as yet not fully interpreted. Require also that these systems aren't in the business of guessing the outcomes of what, by their own lights, are chance events: they never say that *A* without also saying that *A* never had any chance of not coming about. (Lewis 1994, p. 480).

But how probabilistic or 'chance' laws may figure into an ontology guided by Lewis' Humean supervenience that postulates only a vast, contingent mosaic of unconnected events? Lewis explains:

> As before, some systems will be simpler than others. Almost as before, some will be stronger than others: some will say either what will happen or what the chances will be when situations of a certain kind arise, whereas others will fall silent both about the outcomes and about the chances. And further, some will fit the actual course of history better than others. That is, the chance of that course of history will be higher according to some systems than according to others. … [As before] the virtues of simplicity, strength, and fit trade off. The best system is the system that gets the best balance of all three. As before, the laws are those regularities that are theorems of the best system. But now some of the laws are probabilistic. So now we can analyse chance: the chances are what the probabilistic laws of the best system say they are. (Lewis 1994, p. 480).

In Lewis' view, therefore, the laws of nature are those regularities that are theorems of the best theory: the true theory of the world's history of events that best balances simplicity, strength, and likelihood (that is, the probability of the actual course of history, given the theory). If any of the laws are probabilistic, then the chances are whatever these laws say they are. In the last analysis, what is true about the laws and the chances is not determined as a matter of physical necessity. Rather, the objective chances, and the laws that govern them, came to be as they are because of the existence of a pattern in the mosaic of events comprising the history of the world. That pattern — the arrangement of particular facts throughout nature — determines what the chances are. Nothing explains them further.

However this may be, in what follows, we shall not consider specific problems for 'chance' laws within Lewis' framework, neither shall we address the question of whether Lewis' analysis of the notion of chance is consistent, since our evaluation of Lewis' conception of lawhood focuses on the eligibility of his *basic* commitments for a theory of laws — namely, his criteria of 'simplicity', 'strength' and their 'balance' — and these, as already witnessed, remain essentially intact in the extended account that combines the best-system analysis of law and chance together.

Apparently, Lewis' best-system analysis is meant to serve the cause of Humean supervenience. It is of Humean nature because the candidate true systems, among which the best system will emerge, are required to *describe only* the spatiotemporal arrangement of locally intrinsic properties of points or point-sized occupants of points. In other words, the arrangement of qualities in space and time provides the candidate true systems, whereas considerations of simplicity, of strength, and of balance between them do the rest. After all, according to the thesis of Humean supervenience, as described and developed by Lewis, the history of a world's distinct events is all there is to that world. Consequently, in Lewis' conception of lawhood there are no necessary connections in nature, no causal powers, or metaphysically real laws determining the world's events. The world is built out of self-contained particulars that are unconnected and exist independently of each other. Everything else, including laws, supervenes upon them.

In Lewis' best-system theory of lawhood, therefore, laws have no existence other than as parts of the best systematization of the world's history of events. But what if two or more different systematizations are tied for best? Lewis (1986a, p. 124) initially considered as laws only those axioms or theorems that appear in the intersection of *all* the tied best systematizations. In such an occasion, however, there may be few or no axioms in the intersection, thus, leaving us with next to no laws. Later Lewis acknowledges that ''in this unfortunate case there would be no very good deservers of the name of laws'' (1994, p. 479). And he goes on, by concluding: ''But what of it? We haven't the slightest reason to think the case really arises''.

We will see that this is not Lewis' only dismissive response. Independently of the aforementioned issue, however, we do have reasons to think that his conception of the Humean supervenience of laws does not cohere with actual scientific practice. Maudlin (2007), for instance, underlines the fact that physical science has always postulated the existence of laws without suggesting that the laws themselves supervene on or reduce to local matters of particular fact. Scientists do not even attempt to analyse laws in terms of a supervenience condition upon a physical state of affairs. In fact, contemporary practice in physics runs contrary to a Humean predilection of lawhood; for, physicists seek and use laws of nature, not only to *describe* or *summarize* spatiotemporal patterns of physical events, but also in order to conceive the notion of physical possibility in relation to a law. For physicists almost[19] every allowable model of a set of laws is a physically possible *way* that a world 'governed' by such laws can be. Characteristic example is the Einstein field equations of general relativity, normally viewed as a kind of consistency constraint upon physically possible worlds. In general relativity, a given stress-energy tensor that describes the *total* distribution of mass-energy throughout a region is, in general, *compatible* with many different space-time structures. Consequently, all various cosmological models of general relativity are physically possible relativistic worlds satisfying the general relativistic laws and hence allowed by them. Why then could there not be physically possible worlds which agree on their total physical state but disagree on certain lawlike features? This eventuality is certainly open to physical science; there is nothing in the corpus of physics or at its functional practice which dictates that the laws themselves supervene on or reduce to matters of particular (non-nomic) facts. Contrary to Lewis' neo-Humean metaphysical stand, therefore, a reductive analysis of laws is neither assumed by actual science nor endorsed by contemporary scientific practice. Of course, the appeal to the scientific practice does not amount to a conclusive argument against Lewis', essentially Humean, conception with respect to laws. It certainly suggests, however, that such a philosophical account does not provide laws the role that actual scientific practice requires them to undertake.

In Lewis' best-system analysis of lawhood, the standards of simplicity, strength and their balance are supposed to be those that guide us in assessing the credibility of rival hypotheses as to what the nature of laws is. These standards, however, are inherently *vague* and the results of their application depend heavily on the selected interpretation (or definition) of their meaning. For example, it is well known that simplicity is dependent on the language in which a theory is formulated, and so, in order to have unambiguous results, a Lewisian theorist should restrict the scope of the available candidates to languages with simple predicates expressing *natural* properties, namely, properties having space-time points, or perhaps point-sized bits of matter, as instances. Yet, in this case, the difficulty is to provide a non-circular account of 'naturalness', one which does not refer to nomic concepts. In other words, if simple systems, according to Lewis (1986a, p. 124), ''are those that come out formally simple when formulated in terms of perfectly natural properties'', then, any argumentation of the kind that 'natural properties are the ones that figure in laws' involves a vicious circle, as Lewis himself admits.

But even if one is able to provide senses of simplicity and strength that overcome these objections — although doubtful — the most serious difficulty in our opinion that circumscribes Lewis' conception of lawhood is related to the mind-dependence problem of Lewisian laws. For, as long as the standards of simplicity and strength are *relative* to our needs and concerns, the mind-independence of laws of nature may be questionable. Since Lewis (1986a, p. 123) insists on the *rigid* application of standards — i.e., in any possible world, the best system should be selected using the standards we *presently* use in the actual world — Lewisian laws are not mind-dependent in the sense that had the standards been different then so would have the laws of nature. Nonetheless, this rigidification of the application of the presently used standards of evaluation is poorly qualified, because there seems to be no sufficient reason for establishing that they are the most important, or, that these are the only standards that rational beings could possibly use.[20] Lewis' (1994, p. 479) surprising response to this anticipated objection is that ''if nature is kind to us, the problem needn't arise''. Consequently, if fortune favours us, there will be a unique best system that is robustly best, in the sense that it will come out first under any reasonable standards of simplicity and strength. However, no sufficient reason is provided that this hope is actually realised, and in case that it does not, there simply exist no fact of the matter what the laws of nature are. Furthermore, Lewis' optimism about the nature's 'kindness' does not entail that Lewisian laws are mind-independent. In so far as the criteria employed for the selection of the best system are of an *epistemological* nature, invariance under their variation only means that there is an 'intersubjective' agreement about the meaning of the notion of the best system. This fact, however, does not imply that 'best-ness' is a mind-independent feature of the selected system. As Lewis (1994, p. 479) himself admits:

> *if* nature were unkind, and *if* disagreeing rival systems were running neck-and-neck, then lawhood might be a psychological matter, and that would be very peculiar. . . . But I'd blame the trouble on unkind nature, not on the analysis.

Clearly, Lewis' response concedes to a possible psychological element in his best-system account of laws. What is the best possible systematization of nature is not, therefore, an entirely objective matter. Thus, although Humean regularities are mind-independent facts of the world, their lawfulness may be mind-dependent. For us, surely, this is not a feature that may be accepted wholeheartedly.

Of course, the whole issue of Humean supervenience of laws of nature is so wide that a detailed treatment of it would take us far apart of our subject. Nonetheless, we think that the earlier brief exposition clearly shows that, in absence of any compelling arguments that would persuade us that 'Humean laws' satisfy our main intuitions about lawhood, the Humean supervenience of laws falls short of being an attractive or even an applicable view in contemporary science.

## 7 Concluding Remarks

The previously presented arguments establish the existence of non-spatiotemporal, non-supervenient physical relations, weakening, thereby, the metaphysical doctrine of Humean supervenience. For, contrary to Lewis' contentions, contemporary microphysics reveals the existence of external irreducible relations that do not supervene upon a spatiotemporal arrangement of Humean properties. Specifically, *any* relation of quantum entanglement among the parts of a compound system endows the overall system with properties which are neither *reducible to* nor *supervenient upon* any (intrinsic or extrinsic) properties that can possibly be attributed to each of its parts. Hence, due to the all-pervasive phenomenon of quantum entanglement at the microphysical level, the functioning of the physical world cannot just be reduced to that of its constituent parts in conjunction with the spatiotemporal relations among the parts. In this respect, the assumption of ontological reductionism, as expressed in the thesis of Humean supervenience, can no longer be accepted as a reliable precept of the nature of the physical world and its contents.[21] Furthermore, given that certain fundamental physical relations, like the quantum entangled relations, are not supervenient upon a spatiotemporal arrangement of local intrinsic properties, they surely cannot figure in any candidate true systems of Lewis' Humean analysis of lawhood.

The foregoing considerations also illustrate a more general lesson for metaphysics. Although it may seem quite plausible that no physical relation can *fully* determine the qualitative, intrinsic properties (if any) of its relata, the metaphysical doctrine of Humean supervenience implies that a prior spatiotemporal determination of them can fix *any* relation holding among the relata. This alleged possibility has ultimately been shown to be founded on intuitive grounds or a priori philosophical generalization. For, the undeniable existence of non-supervenient relations within fundamental physics induces one to admit that certain relations (on a par with certain qualitative, intrinsic properties) are basic constituents of the world. Weak or strong non-supervenience of specific relations upon non-relational facts clearly indicates that such relations have a physical reality on their own. They should be admitted, therefore, into our natural ontology as genuine irreducible elements. Consequently, a metaphysic of relations of a moderate kind — far from being deemed as paradoxical, as frequently is the case in current philosophical thought — ought to be acknowledged as an indispensable part of our understanding of the natural world at a fundamental level.

**Acknowledgements**

For discussion and comments on previous versions, I thank participants at audiences in the Fifth European Conference of Analytic Philosophy (Lisbon) and Thirteenth International Congress of Logic, Methodology and Philosophy of Science (Beijing).

**Notes**

[1] There is no unanimously accepted definition either of the notion of an intrinsic property, or the conception of a qualitative property. For an introductory survey on this matter, see, Weatherson (2002), especially Section 3.1. Various suggestions for defining intrinsic properties are found, for instance, in Lewis (1986a, pp. 262-66) and, more recently, in Humberstone (1996) and Vallentyne (1997). Compare with Langton and Lewis (1998); Lewis (2001). Here, in an intuitive manner, and for purposes of fixing the terminology to be used in the sequel, a property of an object is considered as being *intrinsic* (and hence *non-relational*) if the object has this property in and of itself, independently of the existence of other objects, and a property is regarded as being *qualitative* if its instantiation does not depend on the existence of any particular individual. For example, the property of owing a particular thing or being the father of a particular person are common examples of non-qualitative, individual relational properties.

[2] A possible world is unified by a natural external relation in Lewis' sense, if and only if, every part of a world bears some such relation to every other part, but no part of one world ever bears any such relation to any part of another world (Lewis 1986b, sec. 1.6).

[3] Lewis' thesis of Humean supervenience has been criticized from the viewpoint of quantum mechanics, partly, by Oppy (2000) and Maudlin (2007).

[4] Characteristic is the following claim of Hume (1975/1748, p. 74) in his "Enquiry Concerning Human Understanding", arguing against the epistemic accessibility or, according to the customary interpretation of Hume, against the reality of necessary connections in nature:

> upon the whole, there appears not, throughout all nature, any one instance of connexion which is conceivable by us. All events seem entirely loose and separate. One event follows another; but we never can observe any tie between them. They seem *conjoined*, but never *connected*. And as we can have no idea of any thing which never appeared to our outward sense or inward sentiment, the necessary conclusion *seems* to be that we have no idea of connexion or power at all, and that these words are absolutely without any meaning, when employed either in philosophical reasonings or common life.

[5] Akin to both principles of separability and Humean supervenience is the thesis of 'local physicalism' or 'particularism' as has been variously called by Teller (1986, 1989), in order to encapsulate the mechanistic worldview of classical physics. According to Teller (1989, p. 213), particularism holds that the basic inhabitants of our world are distinct individuals, possessing non-relational properties, and that all relations between individuals supervene on the non-relational properties of the relata. This means that if a, b, c, … is a set of individual objects, then, any relational properties and physical relations holding among the relata a, b, c, … supervene upon their non-relational properties. It should be underlined that although the doctrines of Humean supervenience and particularism bear strong similarities to each other, they are not conceptually identical. It is not clear, for instance, whether, according to Teller's notion of particularism, spatial or spatiotemporal relations enjoy a privileged status. If particularists treat all relations alike, then spatiotemporal relations ought to supervene upon the non-relational properties of space-time points or of point-objects that occupy space-time points. Hence, in such a circumstance, there appears a difference with the thesis of Humean supervenience, according to which, spatiotemporal relations belong to the subvenient basis, namely, to the set of basic lower-level properties on which upper-level properties supervene. Furthermore, the range of application of each doctrine varies. In contrast to the conception of Humean supervenience, which concerns *all* the properties and relations that do not belong to the subvenient basis, particularism concerns *only* the relations holding among individuals.

[6] It is of no coincidence, in this respect, that Lewis' examples of 'Humean' properties, namely, intrinsic properties that require no more than a spatiotemporal point to be instatntiated, are the values of the electromagnetic and gravitational fields (see also Loewer (1996, p. 102)). On the other hand, Robinson (1989) and, more recently, Oppy (2000) and Butterfield (2006) have argued that vectorial (or, more

generally, tensorial) quantities cannot be 'Humean' properties in Lewis' sense, because vector-valued magnitudes cannot be instantiated at 'isolated' points, as no direction can be associated with 'isolated' points. We need the notion of the 'neighbourhood' of a point in order to define a tangent vector and hence a specific direction. Admittedly, this necessity is explicit in the case of the coordinate-dependent definition of a tangent vector at a point, where in order to define the coordinates of the vector, we necessarily use coordinate variations (differentials) in an infinitesimal area around the point. Even in the case of coordinate-independent definitions — the vector being considered as an equivalence class of curves on the manifold, or as a differential operator acting on differentiable functions of the manifold — there is an implicit necessity in using the notion of the neighbourhood of a point. This being so, we take nonetheless the view that it is not necessary to put much weight on the notion of 'neighbourhood' while investigating the issue of Humean supervenience and its philosophical implications. Thus, from now on, we shall take for granted that tensorial quantities instantiated at space-time points are the most plausible candidates for the properties of the Humean subvenient basis.

[7] Two points of clarification should be noted: First, the symbols " " and "◊" denote the modal operators for necessity and possibility. Thus, " P" should be read "it is necessary that P" and "◊ P" should be read "it is possible that P". Second, the differentiation between 'determinable' and 'determinate' should be understood in the sense of a determinable attribute of a certain kind — such as having spin — and a determinate attribute — such as having spin of, say, 1/2 along a certain direction.

[8] In this work we shall not consider in any detail alternative interpretations to Hilbert-space quantum mechanics as, for instance, Bohm's ontological or causal interpretation.

[9] Lewis touches upon the concept of quantum states, especially the superposition principle of states in quantum mechanics, in the introductory section of his lecture "How Many Lives Has Schrödinger's Cat?", published three years after his untimely death in 2001. Lewis seems to understand a superposed state as a complex of concomitant layers of reality, as a plurality of *coexisting actualities*. Thus, by referring to the example of the benzene ring, he writes: "… a molecule with an objectively indeterminate structure [such as the benzene ring] is really two coexisting molecules, one with one structure and one with the other. (Or at any rate, two things that are molecule-like …)" (Lewis 2004, p. 4). Lewis' rather short account of the quantum superposition concept is sketchy in certain respects, as he himself admits, whereas his conception of 'coexisting actualities' with respect to a superposed state certainly seems ontologically extravagant; instead of one molecule we get a superimposition of probably infinitely many different molecule-like things. This need not be very repugnant for Lewis, in view of his willingness to embrace a kindred doctrine concerning the reality of possible worlds, a doctrine that harks back in his commitment to the actual existence of all those alternative 'realities' entailed by forms of modal reasoning. When it comes to Hilbert-space quantum mechanics, however, we think Lewis' account of the quantum superposition concept is flawed. According to standard interpretational ideas, quantum mechanical superpositions have an interpretation not in terms of what is actual, but rather of what is potentially realizable via a probability distribution of possible values for the various physical quantities of a system out of which *only one* is actualized when the system is interacting with its environment or a pertinent experimental context. Consequently, a superposed state or a general quantum mechanical pure state may be construed in an *ontic* sense, regardless of any operational procedures, as representing a network of potentialities, namely, an intertwined set of potentially possible and not actually coexisting events (cf. e.g. Margenau 1950, pp. 335-37, 452-54; Heisenberg 1974, pp. 16-17; Shimony 1993, ch. 11; Karakostas 2007, pp. 284-87). In the case of the spin-singlet state $W_S$ of Eq. (2), for example, *no* definite spin state can be assigned to either of the particles $S_1$ and $S_2$, since neither of the corresponding unit vectors $|\psi_\pm>_1$ and $|\varphi_\pm>_2$ are eigenstates of $W_S$; the state $W_S$ is an eigenstate of the *total* spin operator $\sigma_1 \otimes I + I \otimes \sigma_2$, which cannot be understood as being composed of *definite individual* spin values of the two single particles. Hence, in the state $W_S$, *no* spin component of either particle $S_1$ or particle $S_2$ exists in an actual form, possessing *occurrent* spin properties. All three spin components of each particle, however, coexist in a potential form and any one component possesses the tendency of being actualized at the expense of the indiscriminacy of the others if the associated particle interacts with an appropriate measuring apparatus. In this respect, the spin-singlet state — as any compound superposed state — represents in essence the entanglement, the inseparable correlation of potentialities, whose content is not exhausted by a catalogue of actual pre-existing values that may be assigned to the spin properties of $S_1$ and $S_2$, separately. As explained in the text, this is an instance, altogether alien to Lewis' ideas, of the peculiar 'holism' of quantum mechanics: definite properties of a compound system do not supervene on properties of its parts and their spatiotemporal relations.

[10] In this connection see Esfeld (2004). Also Rovelli (1996) and Mermin (1998) highlight the significance of correlations as compared to that of correlata.

[11] It is well known that spin-singlet correlations violate Bell's inequalities. We note in this connection the interesting result of Gisin (1991), Popescu and Rohrlich (1992) that for *any* entangled state of a two-

component system there is a proper choice of pairs of observables whose correlations do violate Bell's inequality.

[12] As an attempt to rescue Humean supervenience from the challenge that quantum nonseparability poses, Loewer (1996, p. 104) has suggested that as the ''fundamental space of the world'' be taken to be abstract configuration space, rather than ordinary space-time. In this view — and within the context of Bohm's theory, or more appropriately, a version of it — the quantum state is construed as a genuine field in configuration space, where a 3-dimensional arrangement of an N-particle system corresponds to a single point in 3N dimensions. This, if the property of spin is neglected, otherwise the dimensionality of configuration space increases. The value of the field at any single point in configuration space, then, represents the amplitude of the quantum state at that point. The hope for Humean supervenience is that although fundamental properties are not instantiated at single points or arbitrarily small regions of space-time, once we realize that the ''actual'' or ''fundamental space of the world'' is 3N-dimensional configuration space those properties turn out to be 'local' after all. However, it is not clear whether Lewis' thesis of Humean supervenience retains its substantial content after the move to highly abstract configuration space. The latter is not the concrete spatiotemporal continuum aspired by Lewis. To the extent that the thesis of Humean supervenience enjoys intuitive support, its source no doubt arises from considerations in quality arrangement in space-time. Secondly, even supposing that all presuppositions of Loewer's suggestion do hold, there is no sufficient reason to demand that qualities distributed in configuration space ought to be *intrinsic* to the points of that space. Thirdly, any considerations about the ''fundamental space of the world'' should be grounded in physical science, not in a priori philosophical assumptions. In this respect, contemporary developments in theoretical physics strongly suggest that space-time itself is an approximate, derived notion (e.g., Witten 1996; Butterfield and Isham 2001). On such an account, spatiotemporal relations are no longer fundamental. Hence, the letter of Lewis' formulation of Humean supervenience will be violated. Nonetheless, we shall take no consideration of these possibilities, for even confining ourselves to existing mature, experimentally confirmed physical theories, there exist adequate ground to resist the thesis that Humean supervenience holds in our world.

[13] It is worthy to note that the non-purity of the subsystem states of Eq. (4) arises as a restriction of the overall pure state $W_S$ of the entangled system to the observables pertaining to a component subsystem. Any subsystem state in this situation is exclusively defined at the level of the whole system; there is no individual state for a component subsystem alone. For, there exists no justification in regarding the prescribed reduced states of Eq. (4) as being associated with any specific ensemble of pure states with corresponding eigenvalue probabilities. In this respect, the reference of a reduced state is only of a statistical, epistemic nature. It simply reflects the statistics that may be derived by a series of local measurements performed on a given component subsystem.

[14] Hughston et al. (1993) provide a constructive classification of all discrete ensembles of pure quantum states that correspond to a fixed density operator.

[15] In Lewis' philosophical scheme, the notion of causation should be understood in a broadly neo-Humean way as a *contingent* relation between 'distinct existences'. Causal relationships are, roughly speaking, nothing but patterns which supervene on a point-by-point distribution of properties.

[16] It should be noted that the so-called invariant or state-independent properties — like 'rest-mass', 'charge' and 'spin' — of elementary objects-systems can only characterize a certain class of objects; they can only specify a certain sort of particles, e.g., electrons, protons, neutrons, etc. They are not sufficient, however, for determining a member of the class as an individual object, distinct from other members within the same class, that is, from other objects having the same state-independent properties. Thus, an 'electron', for instance, could not be of the particle-kind of 'electrons' without fixed, state-independent properties of 'mass' and 'charge', but these in no way suffice for distinguishing it from other similar particles or for 'individuating' it in any particular physical situation. For a detailed treatment of this point, see, for example, French and Krause (2006); Castellani (1999).

[17] The seemingly tempting thought of including into the supervenience basis the quantum entangled relations among the related parts of a quantum whole would certainly be inappropriate. For, firstly, such a manoeuvre would run against the very essence of Lewis' thesis of Humean supervenience, envisaging a world of unconnected, independently existing particulars. And, secondly, in such a circumstance any relevant supervenience claim would be rendered trivial. It is essential that in specifying a supervenience basis one should avoid adding *global* properties or relations to this basis. The latter is expected to include, according to Healey (1991, p. 401), only "the qualitative, intrinsic properties and relations of the parts, i.e., the properties and relations that these bear in and of themselves, without regard to any other objects, and irrespective of any further consequences of their bearing these properties for the properties of any wholes they might compose".

[18] See, for instance, Aspect et al. (1982); also the relatively recent result of Tittel et al. (1998).

[19] The qualification concerns models characterized by 'extremely', according to the present state of physical science, counter-intuitive features.
[20] In this connection, see, Armstrong (1983, p. 67) and van Fraasen (1989, p. 46).
[21] It is worth noting that the thesis of Humean supervenience is not identical to physicalism or physicalistic reductionism. It is rather a particularly strong version of it. For, one need not uphold Humean supervenience in order to support a robust physicalist doctrine, in view of which, all contingent matters of fact supervene on physical matters of fact.